\documentclass[aps,prl,twocolumn,showpacs,superscriptaddress,groupedaddress]{revtex4}  
\usepackage{graphicx}  
\usepackage{dcolumn}   
\usepackage{bm}        
\usepackage{amssymb}   
\usepackage{amsmath}
\usepackage{color}
\usepackage[colorlinks=true, pdfstartview=FitV, linkcolor=blue, citecolor=blue, urlcolor=blue]{hyperref} 
\usepackage{epstopdf}

\usepackage{appendix} 

\begin{document}

\preprint{}

\title{Gate-controlled phase switching in a Parametron}
\author{$\check{\mathrm{Z}}$. Nosan}
\affiliation{Institute for Solid State Physics, ETH Zurich, 8093 Zurich, Switzerland}

\author{P. M\"arki}
\affiliation{Institute for Solid State Physics, ETH Zurich, 8093 Zurich, Switzerland}

\author{N. Hauff}
\affiliation{Institute for Solid State Physics, ETH Zurich, 8093 Zurich, Switzerland}

\author{C. Knaut}
\affiliation{Institute for Solid State Physics, ETH Zurich, 8093 Zurich, Switzerland}


\author{A. Eichler}
\affiliation{Institute for Solid State Physics, ETH Zurich, 8093 Zurich, Switzerland}
\date{\today}

\begin{abstract}
The parametron, a resonator-based logic device, is a promising physical platform for emerging computational paradigms. When the parametron is subject to both parametric pumping and external driving, complex phenomena arise that can be harvested for applications. In this paper, we experimentally demonstrate deterministic phase switching of a parametron by applying frequency tuning pulses. To our surprise, we find different regimes of phase switching due to the interplay between a parametric pump and an external drive. We provide full modeling of our device with numerical simulations and find excellent agreement between model and measurements. Our result opens up new possibilities for fast and robust logic operations within large-scale parametron architectures.
\end{abstract}


\maketitle


Many fascinating and useful phenomena arise when the spring constant of a resonator is varied periodically in time. `Degenerate parametric pumping' refers to the important case when the modulation rate is close to twice the natural frequency of the resonator, $f_p \sim 2 f_0$~\cite{Lifshitz_Cross}. As long as the modulation depth $\lambda$ is below a threshold value $\lambda_{th}$, parametric pumping simply decreases or increases the effective damping of the resonator in response to external forcing. This effect is used with great success for quantum-limited signal amplification with superconducting Josephson circuits~\cite{Kuzmin_1983,Yurke_1989,Castellanos_2007,CEichler_2014,Roy_2016}, feedback damping of nanomechanical resonators and trapped particles~\cite{Rugar_1991,Villanueva_2011,Gieseler_2012}, and squeezing of the vacuum noise of laser light~\cite{Caves_1981,Slusher_1985,Breitenbach_1997,Eberle_2010}.

When the modulation depth exceeds the threshold, $\lambda > \lambda_{th}$, the effective damping drops below zero. The system then becomes a parametric phase-locked oscillator, or `parametron', that undergoes large oscillations at $f_p/2$ even in the absence of external forcing~\cite{Lifshitz_Cross,Mahboob_2008,Wilson_2010,Lin_2014}. Due to the periodicity doubling between pump and oscillation, the parametron features two `phase states' that have equal amplitude but differ in phase by $\pi$. In a classical system, the parametron selects one out of the two phases states during a spontaneous time-translation symmetry breaking event, while a quantum system can reside in a superposition of both states. This state duality has been applied to classical computing before the invention of the transistor~\cite{Goto_1959}, and has recently been rediscovered in the context of alternative computational architectures such as neural networks, adiabatic quantum computing and quantum annealing~\cite{Kirkpatrick_1983,Georgescu_2014,Goto_2016,Puri_2017}. Namely, the two phase states can represent opposite values of a single variable, e.g. the polarization states of a single spin (`up and down'). An ensemble of parametrons is envisioned as a simulator to find the ground state of Ising Hamiltonians comprising many spins, or equivalent combinatorial optimization problems from other fields. These problems are NP-hard and thus very challenging to solve on conventional computers. Several exciting physical implementations are currently competing to demonstrate such novel computing paradigms, including optical parametric oscillators, superconducting Josephson circuits, and nanomechanical resonators~\cite{Marandi_2014,Mahboob_2016,Inagaki_2016}.

\begin{figure}
\includegraphics[width=\columnwidth]{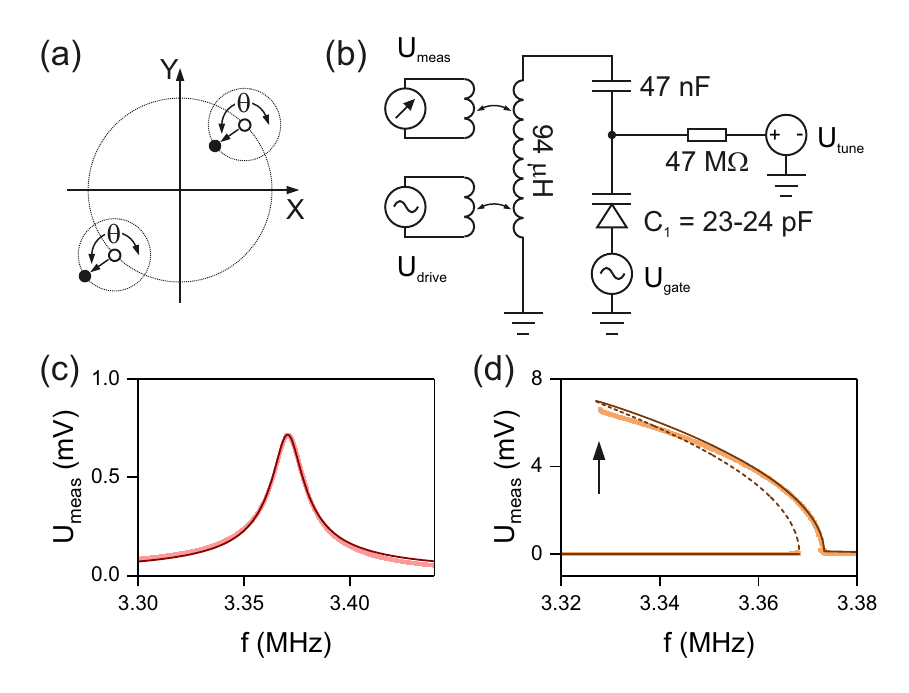}
\caption{\label{fig:Fig1}(a) Phase states of a parametron in the rotating frame (open dots). In-phase amplitude is denotes by $X$, out-of-phase amplitude by $Y$. A resonant force with relative phase $\theta$ can break the symmetry of the parametron and shift the states (solid dots). (b) In our experiment, an electrical resonator is driven and measured inductively. The varactor diode with capacitance $C_1$ is biased with $U_{tune}$ and $U_{gate}$ for DC and AC tuning of $f_0$, respectively. (c) Response to external driving with $U_d = 50$\,mV ($U_p = 0$) and (d) parametric response to $U_p = 5$\,V ($U_d = 0$) with $U_{tune} = 2.2$\,V. Bright dots represent data, fits are shown as lines. The dashed line in (d) indicates an unstable solution branch. From the fits, we obtain $f_0 = 3.37$\,MHz, $Q = 243$, $F_d / U_d = 2.65\times 10^{10}$\,s$^{-2}$, $\alpha = 3.2\times 10^{17}$\,V$^{-2}$s$^{-2}$, and $\eta = 5.8\times 10^{8}$\,V$^{-2}$s$^{-1}$.}
\end{figure}

The symmetry between the two phase states can be broken by applying a force at the oscillation frequency $f_p/2 \sim f_0$ (Fig.~\ref{fig:Fig1}a)~\cite{Ryvkine_2006,Kim_2010,Papariello_2016,Leuch_2016}. Depending on the relative phase $\theta$ of this force, the parametron will favor one of the states, which can be used for various detection and amplification schemes~\cite{Rhoads_2010,Mahboob_2010,Lin_2014,Eichler_2018}. The symmetry breaking has also been found to qualitatively change the bifurcation topology of the parametron, i.e. the way stable and unstable solutions merge and annihilate as a function of $f_p$~\cite{Leuch_2016}. This has surprising consequences for the hysteresis observed when sweeping the pump (at $f_p$) and the force (at $f_p/2$) simultaneously. One of the applications predicted in Ref.~\cite{Leuch_2016} is that the parametron can be switched from one phase state to the other simply by changing the resonance frequency $f_0$ relative to $f_p/2$ with a gate voltage. The `gate-parametron' is effectively a new device with a suite of useful properties.

In this paper, we experimentally demonstrate gate-controlled switching of the phase states of a parametron. With an electrical resonator circuit, we first establish the existence of a double hysteresis due to parametric symmetry breaking, which previously has been found in mechanical resonators at much lower frequencies~\cite{Leuch_2016,Eichler_2018}. We then proceed to demonstrate phase switching with high fidelity and in close agreement with numerical simulations. The achievable switching rate is ultimately limited by the ringdown time $\tau = \frac{Q}{\pi f_0}$ of the resonator (where $Q$ is the quality factor). We find that the switching works down to a timescale of $7 \tau$, where the dynamics is much too fast for the gate-parametron to follow its steady-state response. We believe that the surprisingly fast phase switching stems from the interplay between parametric pumping and external forcing.

\begin{figure}
\includegraphics[width=\columnwidth]{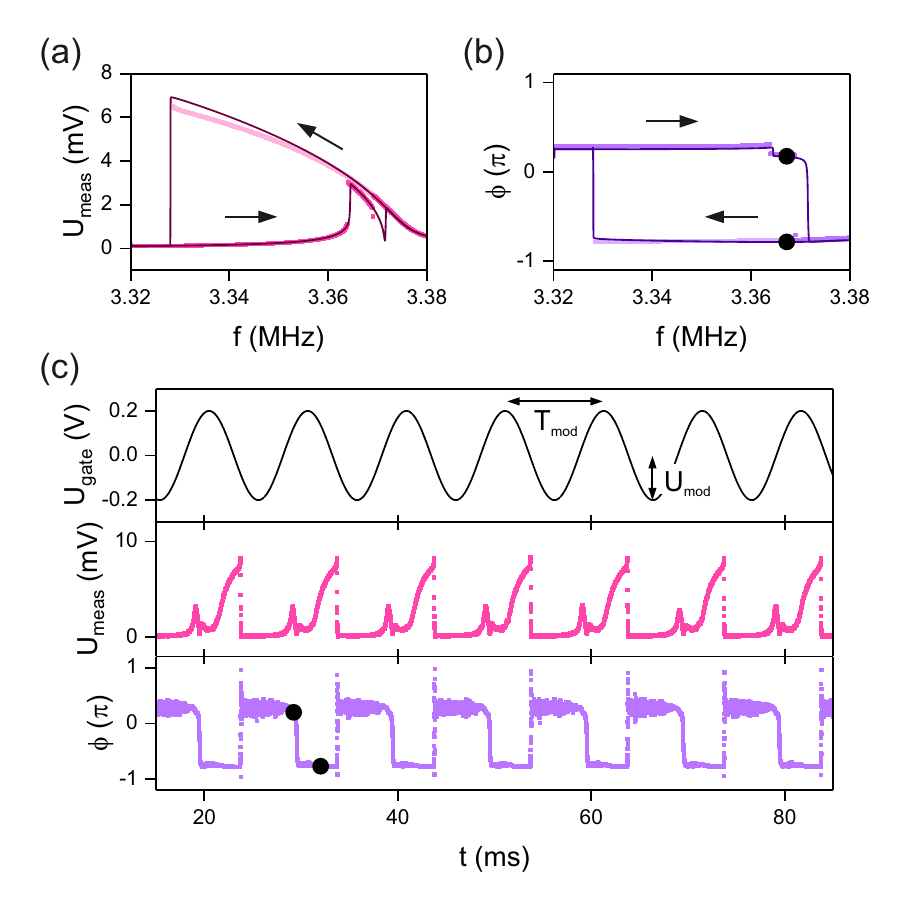}
\caption{\label{fig:Fig2}(a) Amplitude and (b) phase response during sweeps of $f$ with parametric pumping (at $f_p = 2f$) and external driving (at $f_d = f$) applied simultaneously. A characteristic double hysteresis is observed, with two jumps when sweeping from low to high frequency and one jump in the opposite direction. In all graphs, bright squares are measurements and dark lines are simulations. Black dots in (b) indicate the position of the phase states in frequency. $U_{tune} = 2.2$\,V, $U_{gate} = 0$, $U_p = 5$\,V, $U_d = 50$\,mV, $\theta = \pi/4$. (c) Response to a periodic voltage $U_{gate}$ that shifts $f_0$ relative to the fixed $f_{p,d}$. Black dots indicate the approximate positions of the two phase states within one $U_{gate}$ period. $f_p = 2f_d = 6.74$\,MHz, $U_{mod} = 0.2$\,V and $T_{mod} = 10$\,ms.}
\end{figure}

We perform experiments with an electrical circuit whose main elements are a coil and a varactor diode with capacitance $C_1$ (Fig.~\ref{fig:Fig1}b). The precise value of $C_1$ depends on the applied voltage, giving rise to a nonlinearity (see the supplementary material for details and calibration). We use a DC voltage $U_{tune}$ to ensure that the diode is in reverse bias, while rapid changes of the resonance frequency $\omega_0 = 2\pi f_0 \approx \sqrt{1/LC_1}$ are induced through $U_{gate}$. The latter is applied via an operational amplifier (THS4271D) with a nominal unity gain. The low output impedance of this voltage buffer is necessary to preserve the quality factor of the resonator. The resonator is driven and read out inductively with a lock-in amplifier (Zurich Instruments HF2LI).

The equation of motion that governs our system is
\begin{align}
	\ddot{x} + \omega_0^2 \left[1-\lambda\cos(\omega_p t)\right]x &+ \Gamma\dot{x} + \alpha x^3 + \eta x^2\dot{x} \nonumber \\= & F_d\cos(\omega_d t + \theta) \label{eq:EOM}
\end{align}
where dots denote differentiation with respect to time $t$, $x$ is an oscillating voltage, and $\Gamma = \omega_0/Q$ is the damping rate. The varactor diode gives rise to an approximately linear dependence of the spring constant on $x$. Driving the resonator with a voltage amplitude $U_p$ at a rate $f_p = \omega_p/2\pi$ leads to off-resonant oscillations in the circuit that modulate $\omega_0^2$. In eq.~\ref{eq:EOM}, this parametric effect is accounted for by the second term in square brackets with modulation depth $\lambda \propto U_p$. The nonlinearity is also responsible for the Duffing coefficient $\alpha$ and for the nonlinear damping coefficient $\eta$. The resonator can be driven externally with a near-resonant frequency $f_d = \omega_d/2\pi$, an effective amplitude $F_d \propto U_d$, and relative phase $\theta$. For simplicity, we set $x = U_{meas}$, that is, we treat the voltage in our pick-up coil as the effective resonator displacement. The only modification we incur through this step is that the values of $\alpha$ and $\eta$ will be normalized accordingly (see SM for all details).

\begin{figure}
\includegraphics[width=\columnwidth]{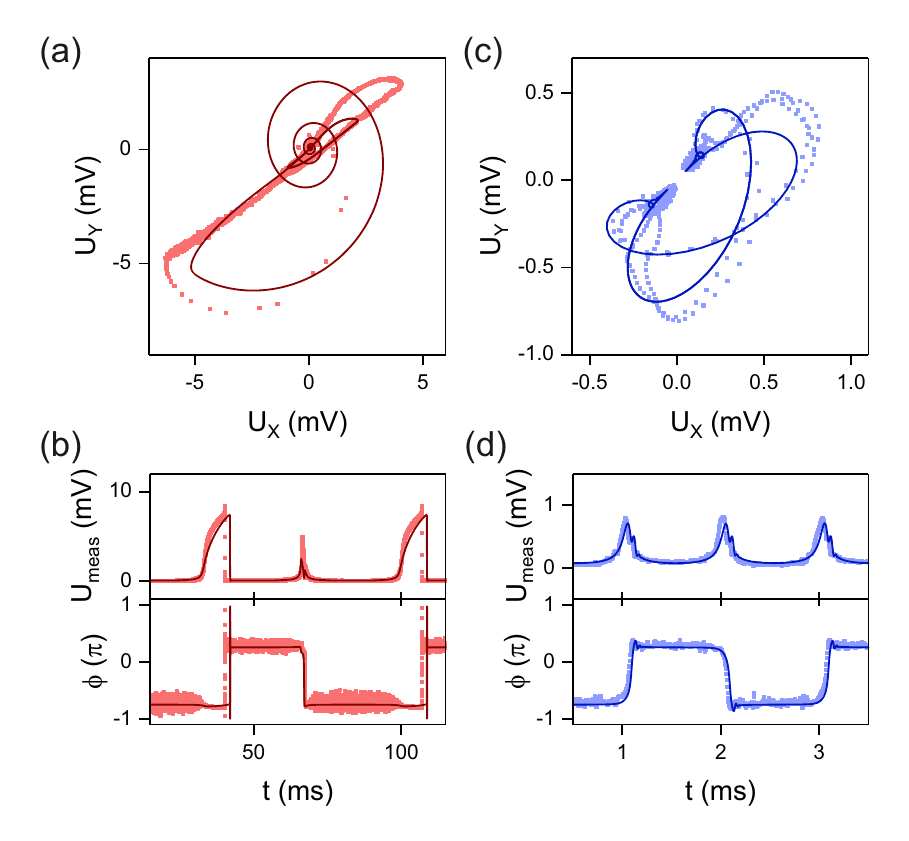}
\caption{\label{fig:Fig3}Periodic response of the parametron to $U_{gate}$ modulations (a) in the rotating frame of the lock-in amplifer (i.e. the in-phase and out-of-phase quadratures $U_X$ and $U_Y$ at $f_p/2$) and (b) in terms of amplitude and phase as a function of time. Bright squares are measurements, solid lines are simulations. The starting phase of the gate voltage modulation is a free parameter in the simulation. $U_{tune} = 2.2$\,V, $U_p = 5$\,V, $U_d = 50$\,mV, $\theta = \pi/4$, $T_{mod} = 67$\,ms, and the pulse has an amplitude of $U_{mod} = 0.45$\,V. In (c)-(d) we show the same for $T_{mod} = 2$\,ms.}
\end{figure}

Under a small external drive and for $\lambda = 0$, the resonator oscillates at $f_d$ with an amplitude that is proportional to $U_d$. From a sweep of the driving frequency, we obtain a Lorentzian response curve that we can use to determine $f_0$, $Q$, as well as to calibrate $F_d / U_d$ (Fig.~\ref{fig:Fig1}c). In the opposite case of purely parametric pumping and $U_d = 0$, a finite response is measured within a certain frequency range when $U_p \geq U_{th}$ (Fig.~\ref{fig:Fig1}d), which allows us to calculate the modulation depth as
\begin{align}\label{eq:threshold}
	\lambda = \frac{U_p \lambda_{th}}{U_{th}} = \frac{U_p}{U_{th}}\frac{2}{Q}.
\end{align}
Beyond $\lambda_{th}$, the device is linearly unstable and enters the nonlinear parametron regime. From a fit to the nonlinear amplitude response, we can extract values for $\alpha$ and $\eta$. Here, the nonlinear damping coefficient $\eta$ is used to model the frequency at which the large-amplitude branch is terminated (see arrow in Fig.~\ref{fig:Fig1}d).


When parametric pumping and external driving are present simultaneously, parametric symmetry breaking occurs~\cite{Leuch_2016}. With $f_p = 2f_d$, this can lead to a complex bifurcation topology and a characteristic double hysteresis in frequency sweeps (Fig.~\ref{fig:Fig2}a and b). Importantly, the external drive causes the parametron to occupy opposite phase states when sweeping the frequency upwards or downwards.

The mechanism that underlies phase switching is surprisingly simple. Far from resonance, the resonator is outside the region of parametric instability and no parametric oscillation takes place. The phase of the resonator is then determined by the external drive alone. Due to the phase difference of the driven harmonic resonator below and above resonance, the external drive imprints opposite phases into the system for the two extreme gate voltages. When the detuning is reduced, the resonator enters the region of parametric instability (either from below or above in frequency) and must ring up to one of the two phase states. In this moment, the phase imprinted upon the parametron by the external drive acts as a bias that deterministically selects one of the two phase states. It was proposed that instead of sweeping $f_{d,p}$, one could vary $f_0$ over time to induce phase state switches~\cite{Leuch_2016}. We present an experimental demonstration of this prediction.

We can change $f_0$ as a function of time by applying a time-varying gate voltage $U_{gate} = U_{mod}\cos(2\pi t/T_{mod})$. In Fig.~\ref{fig:Fig2}c we observe the quasi-static response of the resonator to a modulation of $U_{gate}$ within a period $T_{mod} \gg \tau = 23\,\mu$s. The resonator changes indeed between the two phase states once every $T_{mod}/2$, confirming the possibility of gate-controlled phase switching.

We have tested periodic phase switching with varying speed and found two distinct regimes. In Fig.~\ref{fig:Fig3}a-b the voltage $U_{gate}$ is modulated slowly. Both amplitude and phase follow the response expected from quasi-static frequency sweeps (cf. Fig.~\ref{fig:Fig2}a and b) and the phase space picture in Fig.~\ref{fig:Fig3}a is asymmetric. Upon decreasing the modulation period, we reach a qualitatively different behavior. In Fig.~\ref{fig:Fig3}c-d, the amplitude remains much smaller and the phase space picture is almost symmetric, i.e. the gate-parametron does not follow its steady-state solution. Surprisingly, the resonator still undergoes phase switches. Numerically Runge-Kutta simulations reproduce all of the observed features.

\begin{figure}
\includegraphics[width=\columnwidth]{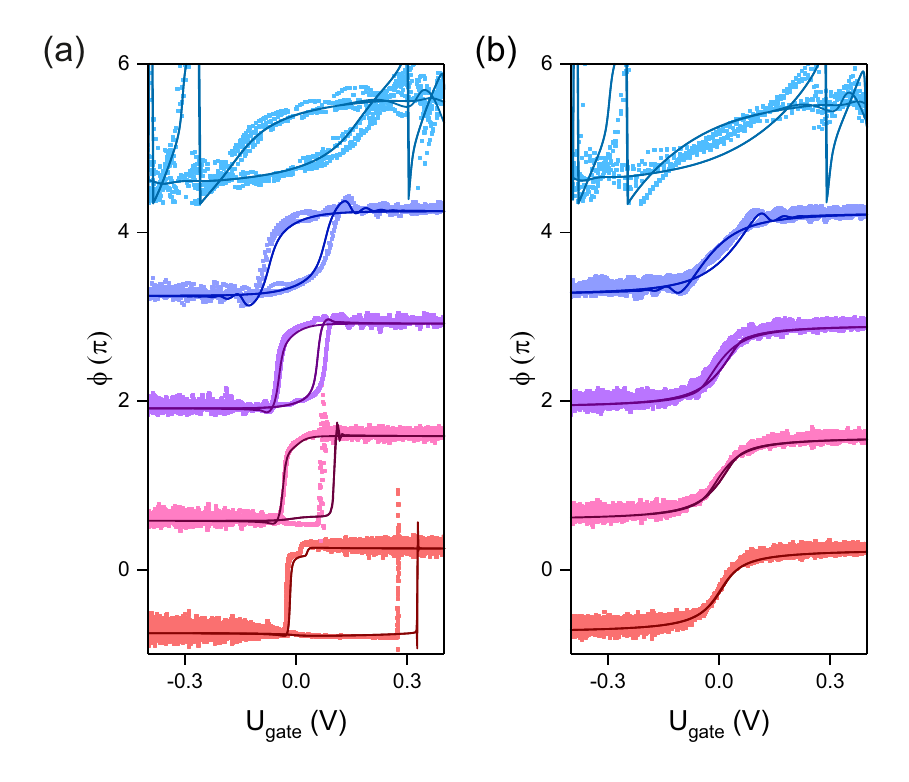}
\caption{\label{fig:Fig4}Hysteresis in the phase response to $U_{gate}$ modulations. Bright squares are measurements, solid lines are simulations, and curves are offset for visibility. (a) Response for $T_{mod} = 67$\,ms, $12.5$\,ms, $6.7$\,ms, $2$\,ms, and $0.33$\,ms from bottom to top, with $U_{tune} = 2.2$\,V, $U_p = 5$\,V, $U_d = 50$\,mV, $U_{mod} = 0.45$\,V, and $\theta = \pi/4$. (b) The same study for $U_p = 0$. Note that the color coding for $T_{mod}$ is consistent with Fig.~\ref{fig:Fig3}.}
\end{figure}

Gate-induced phase switches rely on the formation of a hysteresis. To gain a deeper insight into the different regimes observed in Fig.~\ref{fig:Fig3}, we have investigated the hysteresis of the gate-parametron as a function of $T_{mod}$ in Fig.~\ref{fig:Fig4}a. The large hysteresis observed for slow modulation ($T_{mod} = 67$\,ms, bottom trace) breaks down and reaches a minimum width at $T_{mod} \sim 12.5$\,ms. From there, the hysteresis is found to monotonically increase down to the shortest $T_{mod}$, albeit with a more symmetric shape than for slow modulations. Numerical simulations confirm the experimental results. 

We believe that the hysteresis for short $T_{mod}$ is dominated by the behavior of the underlying harmonic oscillator. If the driving frequency of any resonator is ramped fast enough, the response follows with a delay and a hysteresis develops even in the absence of nonlinearities. In Fig.~\ref{fig:Fig4}b, we have measured the resonator response without parametric pumping. Indeed, we observe that a hysteresis arises for short $T_{mod}$ that resembles the one in Fig.~\ref{fig:Fig4}a.

We conclude that the two regimes of phase switching arise due to the interplay between the parametric pump and the external drive. For slow evolutions, the resonator follows the amplitude and phase dictated by the parametric pump, and the external drive merely acts as a symmetry breaking force. The parametron can directly switch between the slightly asymmetric phase states. For rapid evolutions, it is the external force that dominates the response, and the only visible influence of the parametric drive is a broadening of the hysteresis (compare the upper traces in Fig.~\ref{fig:Fig4}a and b). After the phase flip, the parametron relaxes into the appropriate phase state. Gate-controlled phase state switching is thus still possible in this regime.

As a summary, we have demonstrated the main functionality of the gate-parametron, namely deterministic and gate-controlled phase state switching due to a symmetry-breaking force. More generally, we observed phase switching of the gate-parametron on timescales down to $T_{mod}/2 = 7 \tau$ for a single switch, and we found two distinct regimes of hysteresis formation due to the interplay between the parametric pump and the external drive. Our gate-controlled phase switching technique allows individual control of any number of parametrons sharing one parametric drive at $f_p$ and one external drive at $f_p/2$. It thus offers a simplified architecture for large-scale implementations of parametrons, which may be crucial for novel computation paradigms such as neural networks, adiabatic quantum computing and quantum annealing~\cite{Kirkpatrick_1983,Georgescu_2014,Goto_2016,Puri_2017}. Ongoing and future work will address additional features of the gate-parametron. For instance, modulating the gate voltage of a parametron that is coupled to one or many other parametrons may enable conditional operations, i.e. the phase state that the parametron reaches depends on the states of the surrounding devices and on the gate trajectory. For quantum systems, it will be interesting to study the possibility of rapid generation of quantum superpositions between the phase states of a gate-parametron~\cite{Zhang_2017}, and of entanglement between coupled parametron devices. All of these applications may be implemented in a variety of resonators, ranging from optical parametric oscillators to Josephson junction circuits, nanomechanical resonators, and levitating particles~\cite{Marandi_2014,Mahboob_2016,Inagaki_2016,Gieseler_2012}.

The authors acknowledge fruitful discussions with O. Zilberberg, R. Chitra, and T. Heugel. This work received financial support from the Swiss National Science Foundation (CRSII5\_177198/1) and from a Public Scholarship of the Development, Disability and Maintenance Fund of the Republic of Slovenia (11010-247/2017-12).

\bibliographystyle{apsrev}

\end{document}